Keanu Nakamura

keanunakamura@berkeley.edu

Advanced Tokamak: The Strongly Reversed Central Magnetic Shear Profile

**Abstract:**

This review article will offer a qualitative overview of the strongly reversed shear profile for steady-state operation in tokamaks. For a steady-state reactor to be commercially viable, it is necessary to have a large bootstrap fraction. Currently, there appears great potential in an Advanced Tokamak (AT) regime, namely the hollow current profile (strongly reversed shear). This mode is characterized by high poloidal beta, broad current profiles, strong internal and edge pressure gradients, and relatively good magnetohydrodynamic (MHD) stability against Neoclassical Tearing Modes (NTMs) and ballooning modes. The n=1 and n=2 kink modes, resistive wall modes, and double tearing modes are of concern in the reversed shear profile, and avoidance and/or suppression of these modes is necessary. Although there is a relatively low net plasma current in the reversed shear, the regime appears to have excellent energy confinement properties due to the naturally occurring Internal Transport Barriers (ITBs) caused by the substantial bootstrap currents, and Edge Transport Barriers (ETBs), which can form from ELM-free H-Mode (QH-Mode), to form the Quiescent Double Barrier (QDBs). The reversed shear can be generated by freezing the current profile, through MHD effects or substantial heating and/or current drive during the current ramp up phase, and is sustained by off-axis non-inductive current drive sources, such as the Neutral Beam Current Drive (NBCD), Lower Hybrid Current Drive (LHCD), and Helicon Current Drive (HCD). Experimental results by DIII-D, JT-60U, ASDEX Upgrade, JET, PBX-M, COMPASS-D, and K-STAR, simulation models and codes, such as Lower Hybrid Simulation and STELION, and theoretical reactors, such as ARIES-RS, ARIES-AT and SSTR are referenced.

**Introduction:**

The outline of this paper will be as follows. First, there will be a brief qualitative description of MHD equilibrium in a tokamak since it provides the basis of pulsed, hybrid, and steady-state operation. Second, the bootstrap current and the strongly reversed shear profile will be discussed. The relevant aspects of a hollow current profile will be highlighted, such as MHD instabilities and energy confinement. Lastly, the external non-inductive current drive sources will be reviewed for their coupling efficiencies, penetration, and performance in generating off-axis current.

The tokamak is an axisymmetric torus known for its excellent fusion performance, notably energy confinement. It produces MHD equilibrium by balancing the radial expansion forces, generated by the expanding hot plasma, with toroidal and poloidal fields, and the toroidal forces, generated by the toroidal geometry, with the poloidal fields producing rotational transform [1]. In both cases, the poloidal magnetic fields can be generated externally, by the poloidal field (PF) coils, and internally, by the toroidal plasma current. Depending on whether the plasma current is generated by inductive sources, non-inductive sources, or a mix, the tokamak can be short-pulsed, hybrid, or steady-state. This paper focuses on the last case, where the pulse discharges could theoretically be sustained indefinitely.

The main requirement for a tokamak to operate steady-state is to generate its toroidal current with non-inductive sources. Existing reactors, including ITER, utilize the ohmic transformer, also known as the central solenoid, to generate their inductive current and ohmic heating during the pulsed discharges. The discharge time of these pulsed reactors rely on the capacity of the ohmic transformer's magnetic flux. Steady-state reactors differ in that they will generate their current entirely non-inductively, with radiofrequency (RF) waves, Neutral Beam Injections (NBI), and the naturally occurring bootstrap and diamagnetic current. Since steady-state reactors generally desire low recirculating power for economic purposes, there is a large incentive to maximize the naturally occurring electric currents. In other words, a high bootstrap fraction ($f_{BS}$) is optimal.

The bootstrap current ($I_{BS}$) occurs naturally from local pressure gradients, and is generated through the collisional coupling of trapped and passing particles [2]. Due to the relatively inefficient non-inductive current drive sources, it is estimated that the bootstrap fraction must be greater than 75% for steady-state tokamaks to be economically viable [1]. For reference, ARIES-AT, a theoretical 1000 MWe steady-state reactor, has a bootstrap fraction ($f_{BS}$) > 0.9 [3].

An intuitive solution in increasing the bootstrap fraction is to reduce the total plasma current while maintaining or increasing the bootstrap current. This can be achieved with a hollow current profile (central current density ~0), where most of the plasma current is driven off-axis. Since the safety factor (q) is inversely related to the plasma current ($I_P$), the minimum q ($q_{min}$) will occur off-axis. This differs from positive shear plasmas where the safety factor increases monotonically as a function of radius, and $q_{min}$ is located on-axis ($q_0$). The advanced steady-state scenario is known as the reversed shear since the shear, defined as s = (r/q)dq/dr, reverses sign at the q minimum (figure I) [1].

For the generation and sustainment of hollow current and reversed shear profiles, it is essential to optimize/control the current profile with off-axis external sources. It is necessary to have active control over the current profile ($q_{min}$), especially at high beta, as the plasma undergoes various stages of confinement and bootstrap current profiles, such as current ramp-up, H-Mode transition, and internal transport barrier (ITB) formation [4]. The integration control (I-Control) in JT-60U is used to avoid and stabilize MHD activity in high beta plasmas. The JET and DIII-D controllers, on the other hand, are used primarily for sustaining the advanced regime [5] and for regulating the safety factor profile during current penetration [6], respectively.

The strongly reversed shear scenario, as well as other advanced scenarios, are generated by substantial heating and/or current drive during the current ramp up phase, when the beta is still low, in order to freeze the current profile [7]. The freezing of the current profile can also be done with MHD effects, such as with fishbones, as observed in the ASDEX Upgrade [8], and the presence of mild m/n = 3/2 neoclassical tearing modes, as observed in the DIII-D tokamak [7]. The large off-axis bootstrap current also enhances the reversed shear [9].

**MHD Stability**

In order to achieve a high bootstrap fraction ($f_{BS}$>0.9), steady-state reactors will operate with high beta since the poloidal beta is directly proportional to the bootstrap fraction [10]. The reversed shear scenario is advantageous in that it naturally achieves both high poloidal beta and normalized beta [11]. However, since higher $q_{min}$ values have a negative effect on the beta limit (figure II)[56], and the required beta for sufficient bootstrap currents is high, the beta limit will be approached, and the Troyon no-wall beta limit will likely be exceeded. The beta limit is simply represented as

$$\beta_{\max} = \frac{\beta_N I}{aB} \ [1]$$

where I is the plasma current, a is the minor radius, B is the magnetic field, $\beta_{max}$ is the beta limit, and $\beta_N$ is the normalized beta, calculated to be 0.024 (2.4%) and 0.036 (3.6%) in the absence and presence of a perfectly conducting (ideal) wall, respectively, for the n=1 kink modes [12]. Realistically, reactors will operate with a finitely conductive wall, indicating that the active kink modes will be converted into the Resistive Wall Modes (RWMs), which must be actively stabilized. Additionally, when the n=1 stability limit is exceeded, higher n modes become rapidly unstable, indicating a strict beta limit [13]. It has been found that plasma elongation can linearly increase the beta limit associated with the external ballooning-kink instability; however, the elongation is limited by the n=0 vertical displacement instability, which is converted into the n = 0 RWM due to the resistivity of the walls [1]. It has been found that plasma triangularity can also increase the beta limit [1]. In the conceptual designs, ARIES-RS and ARIES-AT, the elongation ($\kappa$) and triangularity ($\delta$) was increased from $\kappa$=1.7 and $\delta$=0.5 [11] to k=2.2 and $\delta$=0.9 [14]. These changes lead to improvements in stable beta and normalized beta operation–from $\beta = 5\%$ and $\beta_N = 4.8\%$, in ARIES-RS, to $\beta = 9.2\%$ and $\beta_N = 5.4\%$ in ARIES-AT [14]. Experiments in the DIII-D [15] and JET [16] tokamaks have shown that beta limits due to the n=1 ideal kink mode can be increased with broader pressure profiles [12]. H-Mode operation is an effective way of broadening the pressure profile [12], but edge localized modes (ELMs) that occur during standard H-Mode regimes often degrade the ITBs [17]. Fortunately, it is possible to operate in H-Mode ELM-free.

With a relatively low plasma current and a set limit on the normalized beta, it becomes clear that the no wall beta limit will likely be exceeded in a reversed shear plasma. This indicates that the n=1 kink mode becomes unstable and is converted into the n=1 RWM due to the resistivity of the walls. It is important to note, however, that although the resistive wall has no stabilizing effect on the marginal beta, it substantially reduces the RWM growth rates to values comparable to the wall diffusion time: a few ms in present tokamaks and 0.3s in ITER [18]. The slowing of the RWM growth rate is important because it allows for the use of the direct feedback systems. The RWM direct feedback systems use magnetic sensor signals to gain information on the amplitude and phase of the RWM and use these signals to control currents in a set of non axisymmetric feedback coils [12]. Another well known stabilization method is plasma rotation with off-axis Neutral Beam Current Drive (NBCD). It is estimated that the critical rotation frequency for n=1 RWM in the H-Mode equilibria is between 0.07 and 0.08 of the Alven transit frequency [3]. Early experiments by DIII-D [19] and PBX-M [20] have demonstrated stable operation above the no-wall beta limit with sufficient plasma rotation for discharges up to 50 times the resistive magnetic field penetration time of the wall [12]. However, there has been difficulties with sustaining wall stabilization with NBCD plasma rotation for longer pulses as when the no-wall beta limit is approached and exceeded, the plasma response increases substantially [12], strongly damping and slowing the plasma rotation. This effect has been observed in DIII-D [21], NSTX [22], and JET [23, 24]. Fortunately, the discovery of the "error field amplification" effect, also known as "resonant field amplification," has effectively resolved the error fields. As the name implies, error field amplification enhances the detection of small error fields [12], which are corrected by the feedback systems, minimizing the enhanced plasma response. With this method, RWM suppression with plasma rotation has become a routine element in DIII-D experiments [25], and wall stabilization with internal control coils has been sustained for longer than 2.5 seconds [26].

Another MHD instability of concern is the n=2 external kink mode [3]. Although this hasn't been observed experimentally in present reactors, where $\beta_N \leq 3 - 4$, it is predicted to be unstable at higher values of $\beta_N$, and with larger edge pressure gradients [3]. Both are an area of concern for the advanced reversed shear scenario. The latter applies specifically to H-Mode due to the formation of Edge Transport Barriers.

Reversed magnetic shear plasmas are expected to be mostly stable to NTMs [12], unlike positive shear plasmas, generally due to higher $q_{min}$ values. Simplistically, NTM modes, m/n = 2/1 and m/n = 3/2, are avoided when $q_{min} > 2$ [12]. NTMs caused by a loss of bootstrap current from the flattening of the pressure profile (H-Mode) can be avoided/stabilized with continuous ECCD and static resonance magnetic perturbation (RMP) [27]. With continuous ECCD, NTMs are avoided even in the presence of fishbones and sawteeth, and $\beta$ can be increased 20-30% above the onset beta without reappearance of the MHD mode [12].

In ASDEX Upgrade, it has been observed that when $q_{min}$ approaches 2, the m/n = 2/1 mode is observed (figure III)[28]. During the 2/1 fishbone activity, the current profile is locally clamped and the energy confinement is maintained, but at the end, $q_{min}$ drops well below 2 to about 1.7, likely caused by the decoupling of the double tearing modes (DTMs) [28]. DTMs are MHD instabilities that occur specifically in the reversed magnetic shear that can break the magnetic surfaces into a helical structure (magnetic islands) substantially enhancing radial transport [27]. The fast reconnection phase during the DTM is also found to cause explosive bursts [29-31], which are related to the major disruptions in the JT-60U experiments.

It has been found, however, that when central ECRH was applied prior to the expected onset, DTMs did not appear, and the DTMs disappeared when ECRH was applied [28]. Since the distance between the two rational surfaces and their differential rotation has not changed, the stabilizing effect of the ECRH is believed to be caused by the increased electron temperature [28].

Concerning the ballooning modes, the combination of reversed central magnetic shear and the high-confinement mode (H-Mode), enables access to the second stability regime in the core plasma [32], where much higher beta values can be achieved [33]. The mitigation of the ballooning mode with the second stability regime is a prerequisite for the formation of ITBs as it allows for the large pressure gradients at the location of the ITBs [33].

**Energy Confinement**

Since the total plasma current is reduced with hollow current profiles, it would seem, from the H-Mode confinement scaling laws (equation 2), that the energy confinement would be worsened. The IPB98 scaling law is given by

$$\tau_e^{IPB98} = 0.145 M^{0.19} I^{0.93} n^{0.41} B^{0.2} k^{0.78} R^{1.39} a^{0.58} P^{-0.69} \text{[eq 2]}$$

Where I, the plasma current, has an almost linear correlation with confinement time [34]. However, this proves not to be the case as the strongly reversed magnetic shear is associated with the formation of Internal Transport Barriers (ITBs).

ITBs are characterized by regions of steep pressure gradients near the plasma core, where there is reduced radial transport (figure IV)[35] as well as reduced ion and electron thermal conductivities (figure V)[33]. ITBs also increase the bootstrap current as their hollow current profile closely aligns with the bootstrap current [1-2, 33]. ITBs are generally formed by off-axis neutral beam injections and/or lower hybrid current drive during the current rise phase [36], which directly corresponds to a high central safety factor $q_0 > 1$ and a non-monotonic q profile (reversed shear).

It is important to note, however, that ITBs can cause the accumulation of impurities, which can increase transport [32,37] and worsen performance due to fuel dilution [38]. Additionally, the local pressure gradients from ITBs can further restrict the maximum achievable $\beta_N$ ($\beta_N \sim 2$) [12]. JT-60U, in the reversed-shear scenario, overcame the low normalized beta limit ($\beta_N < 2.2$) generated by the strong ITBs with very high safety factors (q95~9) to attain a high enough poloidal beta ($\beta_p < 3.1$) for a high bootstrap fraction ($f_{BS} \sim 0.8$) [7]. Nonetheless, strong ITBs with very steep pressure gradients are not optimal as they often lead to MHD instabilities and disruptions [38]. In early Alcator C-Mod experiments, strong ITB formations caused a continuous peaking of the electron density (up to $1 \cdot '10^{21} m^{-3}$) eventually leading to the collapse of the ITB [39]. It was discovered, however, that by adding a modest amount of on-axis heating power (Ion Cyclotron Resonance Heating), the peaking of the electron density can be prevented and the transient ITBs can be sustained for longer periods of time [39] (up to ten energy confinement times) [36].

Along with ITBs, tokamaks with strongly reversed shear plasmas will likely operate in the advanced high-confinement mode (H-Mode). The H-Mode corresponds to enhanced energy confinement approximately double the low confinement (L-Mode). The reduced transport in H-Mode can be attributed to the edge transport barriers (ETBs), which reduce transport at the edge of the plasma. Due to the increased confinement, H-Mode corresponds to the buildup of impurities and edge density, consequently leading to the excitement of Edge Localized Modes (ELMs), which can act somewhat positively as a pressure and impurity relief valve and/or negatively by eroding the divertor plates [40], increasing transport, and reducing stable beta limits (induces MHD activity) depending on the ELM type. In the case of reversed shear profiles, it is essential to prevent the activity of ELMs as they can severely degrade the ITBs [17]. In other words, it is necessary to achieve an ELM-free quiescent H-Mode (QH-Mode) for the simultaneous formation of ITBs and ETBs, known formally as the quiescent double barrier (QDB).

In DIII-D, the QH-Mode is achieved with neutral beam injections (~2.5MW [41]) in the direction opposite to the plasma current in addition to cryopumping to reduce the density [40]. The density control is possible due to the presence of a benign edge MHD oscillation (the edge harmonic oscillation) [40], which enhances edge particle transport while maintaining near neoclassical levels of transport.

In DIII-D, QDBs have been maintained for periods >25 energy confinement times (>3.5s) [40]. Although the durations are relatively short in regards to steady-state operation, it is predicted that there are no known plasma physics limitations that would prevent the QDBs from being sustained indefinitely [40]. In DIII-D experiments, QDBs were limited by the choice of plasma current flat-top and neutral beam pulse length.

**Non Inductive Current Drive:**

Next, the available external non-inductive current drive sources for the strongly reversed shear profile will be analyzed due to the unlikelihood of a full bootstrap current ($f_{BS} = 1$) in a steady-state reactor. To induce a hollow current profile, most of the external current will be driven off-axis (r/a>0.6 [42]). Second, external sources with higher coupling efficiencies are generally more favorable as they lead to a lower net circulating power, and thus, higher bootstrap fractions and fusion gain factors.

**Radiofrequency (RF) Waves**

Radiofrequency (RF) waves are capable of driving large fractions of the plasma current non-inductively. They launch waves that propagate in one direction around the torus in a collisionless damping mechanism known as Landau Damping [1]. Potential sources include Electron Cyclotron Current Drive (ECCD), Lower Hybrid Current Drive (LHCD), and the Helicon Current Drive (HCD). The RF sources that will generally be driving the largest fractions of off axis non-inductive current are LHCD and HCD, and thus, they will be the main focus of this section. ECCD has very high precision, making it ideal for off-axis current profile control and for the avoidance/stabilization of MHD modes (NTMs and DTMs), but its very low efficiencies (kA/MW) result in it playing a minor role in the total current generated [43].

**Lower Hybrid Current Drive**

LHCD currently presents the greatest potential for off-axis current as it has the highest coupling efficiencies and naturally peaks strongly off-axis. In present experiments, off-axis LHCD is oftentimes used to induce reversed shear plasmas and internal transport barriers [32]. In the JT-60U device, it has been demonstrated that LHCD near the ITB location (near $q_{min}$ [7]) can expand the ITB radius to maintain very high confinement, ($H_{98(y,2)} = 1.4$) [7,44]. In COMPASS-D, LHCD at relatively low power (60-70 kW) demonstrated a strong stabilizing effect on the m/n = 2/1 Neoclassical Tearing Mode [45].

LHCD has strong off-axis single pass damping in the outer radius at normalized radius r/a ~ 0.6-0.8 (figure VI)[22, 46] . However, as observed in JET [42], Frascati Tokamak Upgrade (FTU) [47], and Tore Supra [42], LHCD has difficulty penetrating beyond the edge of the plasma at higher (operational) plasma densities and at weaker toroidal magnetic fields. It is predicted that the density and temperature fluctuations as well as strong parametric-instability (PI) effects [47] are the main factors that cause strong damping at the plasma boundary.

When the density is increased from $n_e = 3.5 \cdot 10^{19} m^{-3}$ to $n_e = 1 \cdot 10^{19} m^{-3}$ during L-Mode discharges in JET, the power deposition is shifted substantially off-axis, from normalized radius r/a=0.5 to r/a = 0.9 (figure VII)[42]. In FTU, densities of $n_{e(0.8)} = 0.4 \cdot 10^{20} m^{-3}$ are considered to be the upper limit for efficient LHCD operation at r/a=0.8. This is problematic considering ITER requires densities of $n_e = 0.7 - 1 \cdot 10^{20} m^{-3}$ in the off-axis location (r/a = 0.8) [47].

With increased edge temperatures, it has been found that the strong PI effects are mitigated, and that the LH waves can penetrate into the plasma more effectively at the optimal location (r/a=0.6-0.8). LHCD effects are considerably increased at r/a = 0.8 at densities ($n_{e(0.8)} = 0.4 \cdot 10^{20} m^{-3}$), and LHCD effects have even been observed in FTU at $n_{e(0.8)} = 0.85 \cdot 10^{20} m^{-3}$ [47].

High peripheral electron temperatures in the FTU have been achieved by using a lithium coated vessel interior, and optimized gas fueling and plasma shapes [47]. These led to reduced particle flux and lower radiative losses in the scrape off layer (SOL) [48]. Local heating by ECRH is also highly effective.

**Helicon Current Drive**

Similar to the LHCD, the Helicon Current Drive (HCD) method generates effective off-axis current, but differs in that it naturally has strong damping near the mid-radius (figure VIII)[18]. The helicons are fast waves with high harmonics of the ion cyclotron frequency, but below the slow LH frequency [18, 49]. Due to the quasi-electromagnetic polarization of the helicons, with weaker RF electric field in the direction of the static magnetic field, the wave can reach the core of the plasma with high electron beta before being strongly damped [49]. The ray's trajectory is spiral rather than directly through the center, indicating strong damping and current deposition off-axis (figure IX)[18].

The HCD system has been primarily tested in the DIII-D tokamak. They have produced H-Mode target plasmas with many discharges of electron beta $\beta_e \geq 0.02$ (figure X) and full first pass absorption at the mid-radius region [18, 49]. The HCD is clearly complementary with the LHCD system. A distinct disadvantage of HCD, however, is that the high harmonic fast waves are susceptible to strong energetic alpha particle damping, which reduces absorption by electrons [3].

Although less developed than the other RF sources [3], the HCD shows good prospects due to its comparatively high efficiencies of ~60 kA/MW in the DIII-D tokamak [18, 50]. Calculations from the STELION full wave code show that efficiencies as high as 164 kA/MW could be obtained [50]. The ECCD drives one quarter of the HCD in the same discharge at ~15 kA/MW and 19 kA/MW at lower (optimal) densities [18]. However, LHCD efficiencies remain the highest: in the optimal location just below the high field side (HFS) midplane, DIII-D shot 133103 achieved 210 kA/MW [46]. In the Lower Hybrid Simulation (LHS) code for the K-STAR tokamak, LHCD efficiencies are calculated to be greater than 400 kA/MW [51].

**Neutral Beam Current Drive (NBCD )**

The Neutral Beam Current Drive (NBCD) method can drive a current in the plasma as the fast ions circulate toroidally around the torus, and are not entirely shielded by the electrons [43]. The Neutral Beam Current Drive (NBCD) is particularly well suited for off-axis current drive, and is advantageous over the RF methods, as its efficiency remains high at larger deposition radii, as the higher fraction at a larger radius reduces the electron shielding current [43]. The DIII-D NBCDs peaks near the mid-radius, which is closer to the axis than LHCD and HCD. By modifying the NBI lines to allow vertical steering, the NBCDs peaked significantly more off-axis, from r/a=0 to r/a=0.45 (figure XI)[52]. The alignment of the neutral beam injections in relation to the magnetic field pitch has shown to have a substantial effect on the efficiency and off-axis peaking location. In DIII-D, it has been found that the injections in the positive, or parallel, direction to the toroidal field (counter clockwise from looking at the top of the tokamak) has a current magnitude 40% higher (figure XII)[52]. This favorable alignment is expected to have up to a 20% effect in ITER [52]. In addition to driving off-axis current, NBCD is used extensively in Advanced Tokamak (AT) experiments to induce ITBs and reversed shear profiles with counter-current injections, and to generate plasma rotation to suppress RWM growth.

For current drive, beam energies of 1000 keV - 2000 keV will be required, which can only be supplied by the negative neutral beam injection (NNBI) system [53-54]. At these energies, the neutralization efficiencies for NNBI remain consistent at ~55%, while the positive neutral beam injections (PNBI) reduce drastically to ~0% (figure XIII)[43]. The off-axis NBCD in DIII-D has an efficiency of ~26 kA/MW [18] and has driven 100% non-inductive current drive for longer than 0.5 seconds with 16 MW of supplied power [55-56].

Although the neutralization efficiency of the NNBI system is relatively high at high beam energies, there is a great incentive to increase the efficiencies for higher fusion gain and economic purposes. First, the numerous power losses in the system can be reduced by some factor, particularly those associated with non neutralized fractions. The power losses include stripping losses in the accelerator (NNBI), geometric beam transmission losses, and reionization of a fraction of the beam [43]. For ITER, these losses from NBI are expected to amount to ~45% [57]. It is hypothesized that typical neutralization efficiencies of ~55% can be increased to greater than 70%, which is a goal for the DEMO design [57].

Hopf et al. outlines two advanced neutralizer concepts for minimizing losses specifically from non neutralized fractions: photoneutralization and plasma neutralization [57]. The neutralization of the negative energetic ions by photodetachment (photoneutralization) has a high theoretical neutralization efficiency of >80%, and a wall plug efficiency of >60% [44]. This method is currently under R&D, however, and has not yet been implemented in a fusion reactor. The plasma neutralization method, where the ionized dense neutralizer gas presents large cross sections for the neutralizing ions ($D^- \rightarrow D^0$) and smaller cross sections for the reionizing atoms ($D^0 \rightarrow D^+$), the neutralization efficiencies could reach as high as 80% for an electron line density of $n_e = 2 \cdot 10^{19}$ [44]. For an ITER plasma neutralizer of 3 meters, this corresponds to density of $n_e = 7 \cdot 10^{18}$. For a more thorough and detailed explanation on the operations of photoneutralization and plasma neutralization, readers can refer to [43], [57], and [44].

Additionally, off-axis NBCD can provide plasma rotation, which can suppress the growth of the resistive wall mode (RWM). Since the torque input per beam power increases with decreasing beam energy, beam energies will range as low as 35 keV [53] - 100 keV [58] with power requirements of 10 MW [58] - 76 MW [53]. The plasma rotation with NBCD has also been found to help induce the formation of ITBs in reversed shear discharges in the JT-60U device [9].

**Conclusion**

Due to the low efficiencies of non-inductive current drive methods, tokamaks should rely, as much as practically possible, on the bootstrap current. This can be accomplished with the hollow current profile, or the strongly reversed shear scenario. The net plasma current is reduced, the bootstrap fraction is increased, the poloidal beta is raised, MHD stability against ballooning modes NTMs is achieved, and strong pressure gradients appear (Internal Transport Barriers). The high beta that's required for high bootstrap fractions will likely exceed the Troyon no wall beta limit, inducing internal and external kink modes which are converted into the resistive wall modes. The beta limit can be increased with a nearby conducting wall, and increased plasma triangulation and elongation. The RWMs will have to be stabilized with plasma rotation from off-axis NBCD and/or direct feedback with non axisymmetric feedback coils. The former requires resonant field amplification at high beta due to an increasing plasma response resulting in strong damping and slowing of the plasma rotation. The strongly reversed shear is associated with excellent energy confinement. Although the total plasma current is reduced, the internal transport barriers that appear near the core and the transport barriers near the edge from H-Mode operation greatly reduce transport to near neoclassical levels.

Since steady-state reactors will most likely not have a bootstrap fraction of 1, external non-inductive current drive sources will be required. The three primary sources that will drive the majority of the external plasma current have been described: the Lower Hybrid Current Drive, Helicon Current Drive, and the Neutral Beam Current Drive. The LHCD is optimal for the reversed shear scenario for multiple reasons. It peaks strongly off-axis (r/a=0.6-0.8) and has the highest efficiencies. Although it has difficulty maintaining strong damping in the optimal location at high densities, high peripheral temperatures have proven successful in reducing the strong PI effects. The Helicon Current Drive penetrates more easily into the plasma core, and peaks closer to the mid-radius, which is complementary with LHCD. However, it currently lacks the extensive research that LHCD, ECCD, and NBCD systems have. The Neutral Beam Current Drive peaks closer to the axis than HCD, but remains fairly effective at various deposition radii. It can provide strong off-axis current that is sufficient for the advanced scenario, as well as inducing strong plasma rotation for MHD stabilization, and ITB formation.

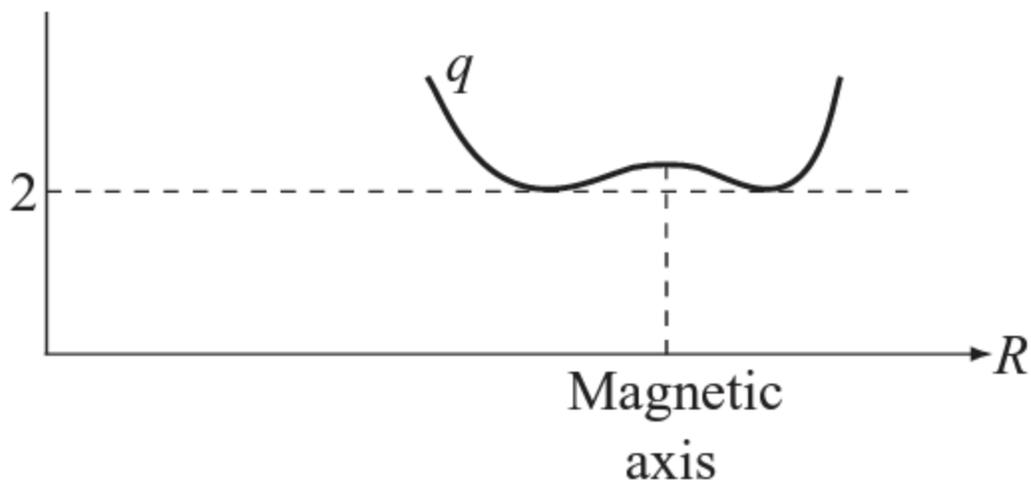

**Fig I** [1] Off-axis $q_{min}$ in reversed shear in DIII-D tokamak

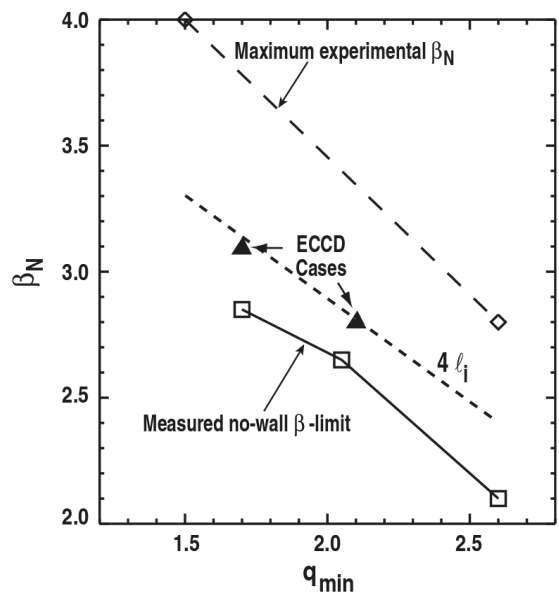

**Fig II** [56] The no-wall beta limit and maximum normalized beta as functions of $q_{min}$. Experimentally observed discharges in the DIII-D tokamak

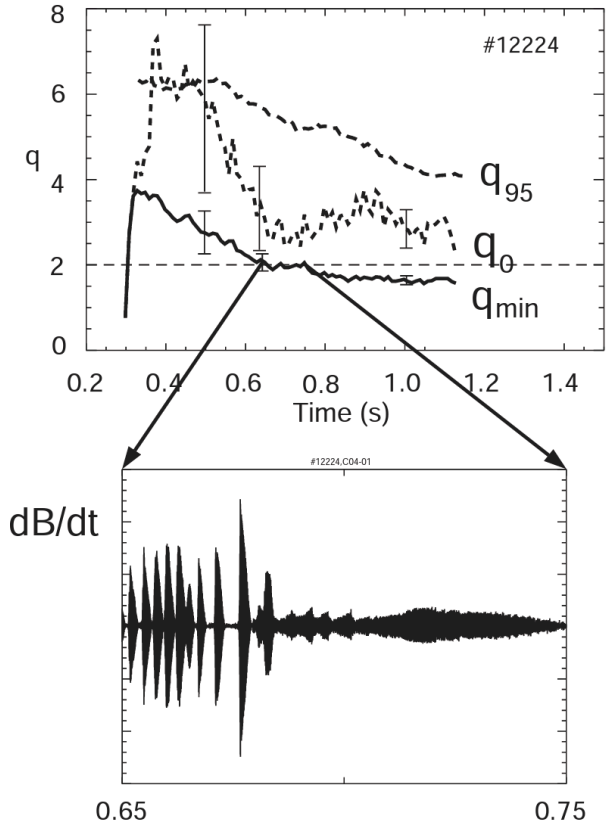

**fig III** [28] NTM mode m/n = 2/1 is observed when $q_{min}$ approaches 2. Between 0.65 and 0.68 seconds, 2/1 fishbones are observed, and afterwards, a 2/1 continuous mode is observed

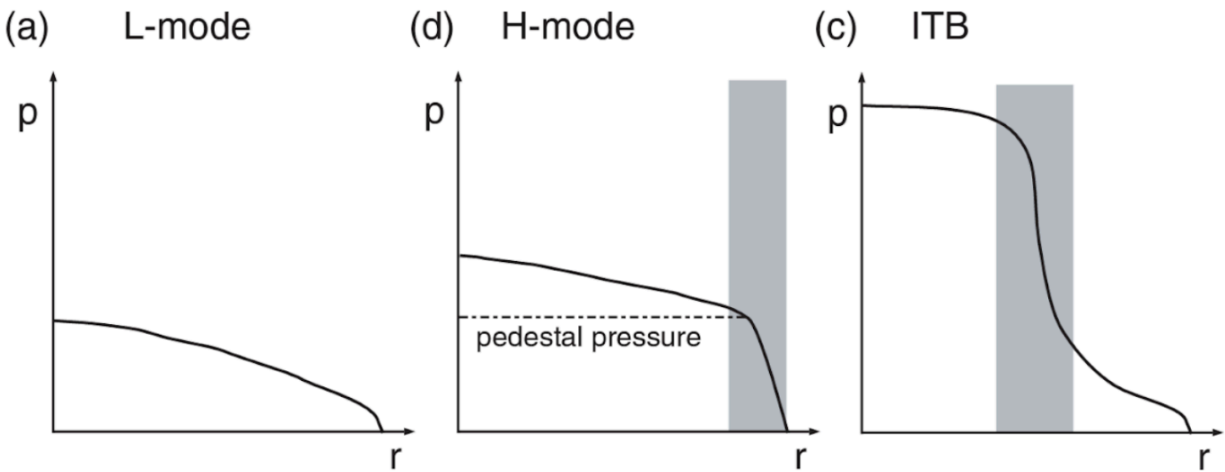

**Fig IV** [35] Pressure profiles of operating regimes. The shaded areas indicate regions of reduced radial transport. ITB occurs in the plasma core, Edge Transport Barriers (ETBs) in H-Mode occur at the edge

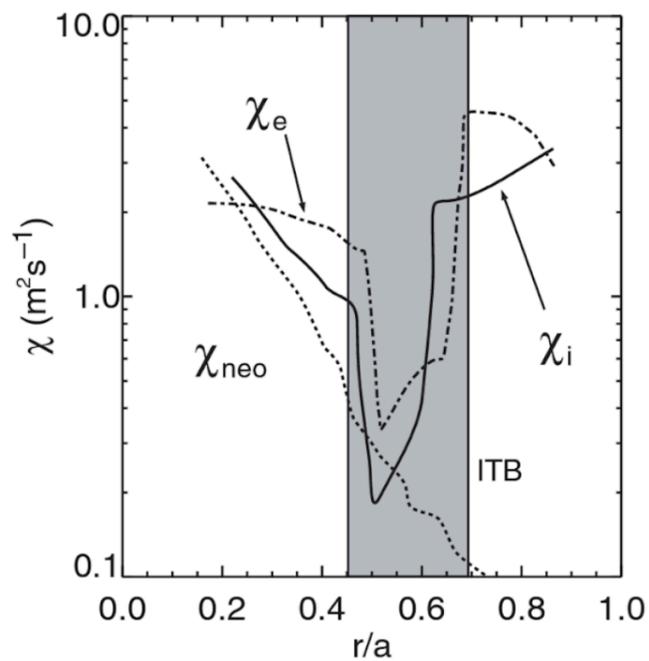

**Fig V** [35] Reduction of ion and electron thermal conductivities associated with ITBs

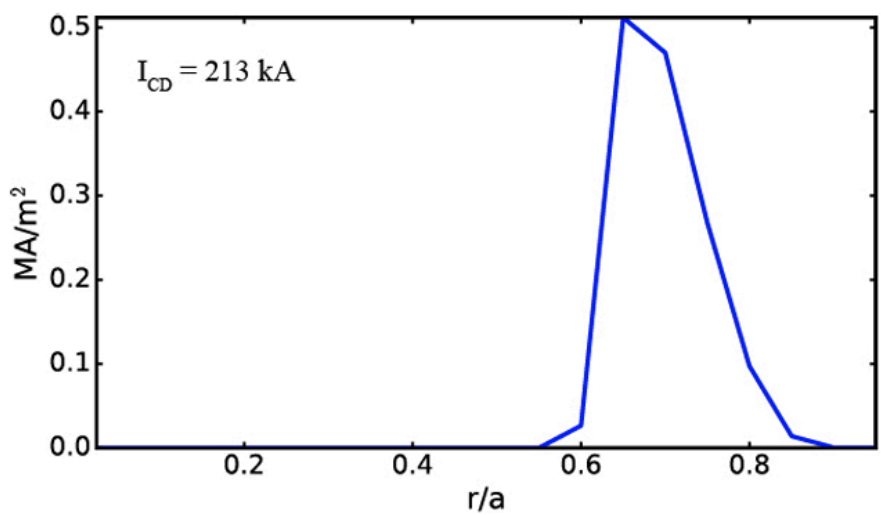

**Fig VI** [46] LHCD strong single pass damping at r/a ~ 0.6-0.8

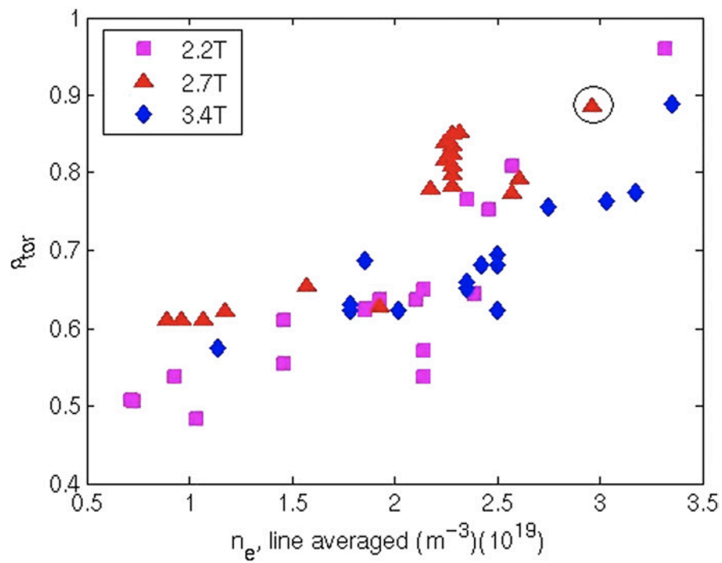

**Fig VII** [42] Recorded LHCD power deposition radii in JET as a function of density using LH power modulation technique

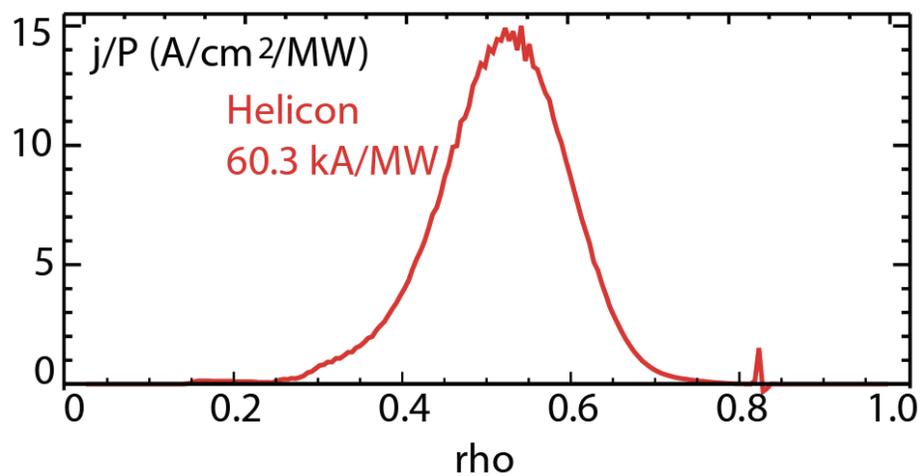

**Fig VIII** [18] Helicon Current Drive efficiency in DIII-D at different radii. It peaks near the mid-radius (r/a ~ 0.5)

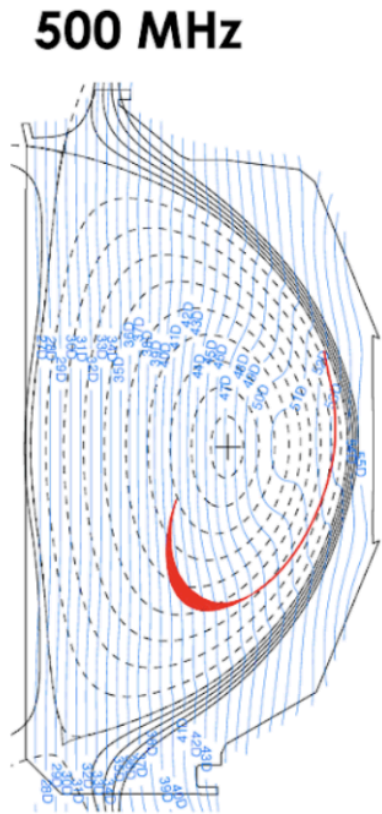

**Fig IX** [18] Spiral trajectory of helicons in DIII-D tokamak

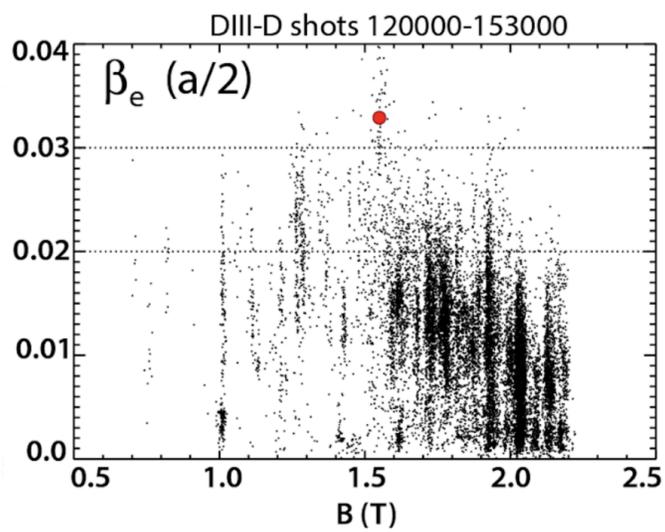

**Fig X**[18] Electron beta as functions of toroidal field in DIII-D tokamak

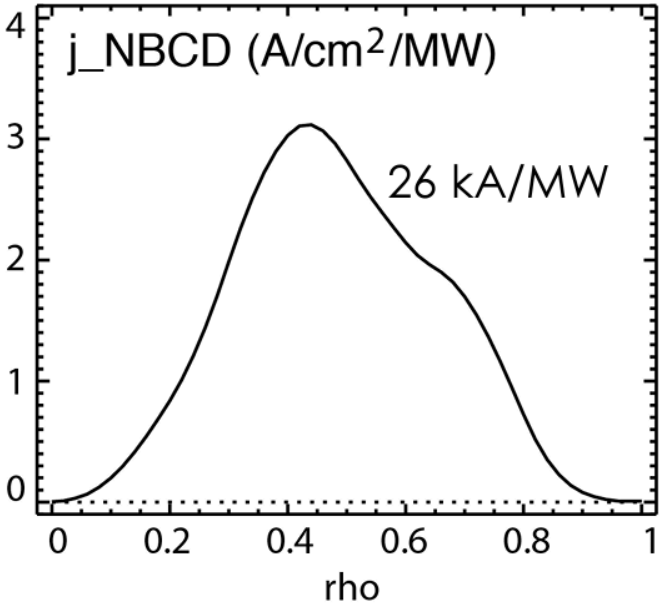

**Fig XI** [18] Neutral Beam Current Drive efficiency in DIII-D at different radii. It peaks at r/a ~ 0.45

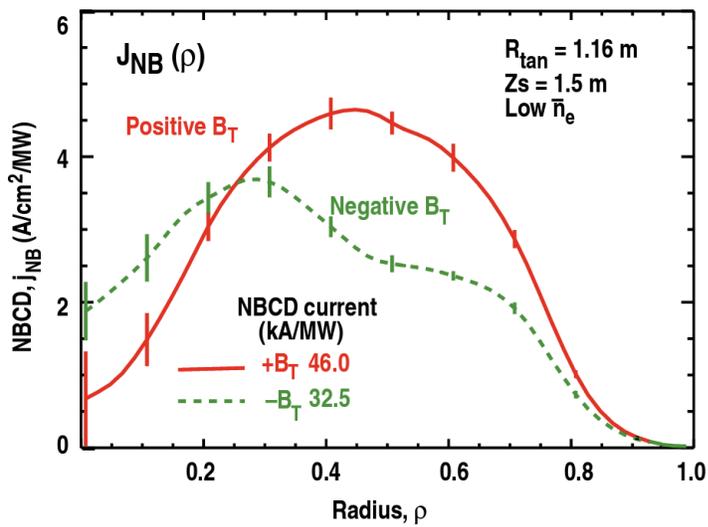

**Fig XII** [52] Positive (favorable) and negative (unfavorable) toroidal field directions for NBCD in DIII-D. The favorable alignment peaks more strongly off-axis, near r/a=0.45

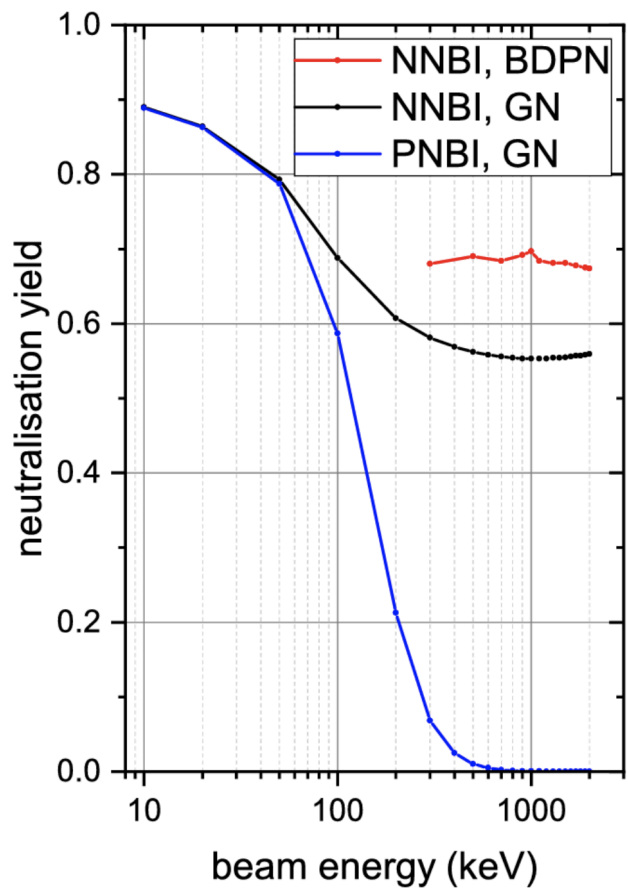

**Fig XIII** [43] Neutralization efficiencies of positive and negative neutral beam injections as functions of beam energies. GN: gas neutralizer. BDPN represents the advanced plasma neutralization method